\documentclass[twocolumn,aps,nofootinbib]{revtex4}
\usepackage[letterpaper, margin=1in]{geometry}
\usepackage[pdfpagelayout=SinglePage]{hyperref}
\usepackage{graphicx}
\usepackage{epstopdf}
\usepackage{latexsym}
\usepackage{amsmath}
\usepackage{amssymb}
\usepackage{float}
\usepackage{subfig}
\usepackage{color}
\usepackage{subcaption}
\usepackage{mathrsfs}
\usepackage{lipsum}
\renewcommand{\Re}{\operatorname{Re}}
\renewcommand{\Im}{\operatorname{Im}}
\usepackage{float}
\usepackage{algorithmic}
\usepackage{algorithm}
\usepackage{amsmath}
\usepackage{enumerate}
\usepackage{xcolor}
\usepackage{makecell}

\begin{document}

 \newcommand{\bq}{\begin{equation}}
 \newcommand{\eq}{\end{equation}}
 \newcommand{\bqn}{\begin{eqnarray}}
 \newcommand{\eqn}{\end{eqnarray}}
 \newcommand{\nb}{\nonumber}
 \newcommand{\lb}{\label}
 \newcommand{\be}{\begin{equation}}
\newcommand{\en}{\end{equation}}
\newcommand{\PRL}{Phys. Rev. Lett.}
\newcommand{\PL}{Phys. Lett.}
\newcommand{\PR}{Phys. Rev.}
\newcommand{\CQG}{Class. Quantum Grav.}
\title {Quasinormal  Modes of Bardeen Black Hole in 5-dimensional Gauss-Bonnet Gravity  }
\author {Feng-Miao Ge}\email{gefengmiao@cug.edu.cn}
\author{Bing-Yuan Wei}\email{1256444040@qq.com}
\affiliation{School of Mathematics and Physics, China University of Geosciences, Wuhan 430074, Hubei, China}

\begin{abstract}

This study addressed the scalar field quasinormal ringing behavior of black  holes. We investigated scalar field perturbations in Bardeen black hole spacetime in 5-dimensional Einstein-Gauss-Bonnet (EGB) gravity. Using the  6th order Padé approximation and the finite-difference method, we computed the frequency of quasinormal modes (QNMs) in the spacetime background. The calculations demonstrated that the real part of the QNMs  $\omega$ increased, whereas  the imaginary part decreased with increase in the magnetic charge parameter $Q$ of the Bardeen black hole for a fixed Gauss-Bonnet parameter $\alpha$. This was also valid when $Q$ was fixed and $\alpha$ increased; where in the real part of the QNMs  increased and the absolute value of the imaginary part decreased. However, the change in the latter case was more significant than that in the former; thus, the frequency of eigenvibration of this black hole background under the scalar field perturbation increased and the decay of eigenvibration decreased with increase in $\alpha$ or $Q^2$.
Moreover,  this result shows that the effect of $\alpha$ on the intrinsic vibration of this black hole was greater than that of $Q $.

\end{abstract}


\maketitle

\section{Introduction}
\renewcommand{\theequation}{1.\arabic{equation}} \setcounter{equation}{0}


The singularity problem remains an open puzzle in general relativity. This is because the classical model describing gravity typically results  in the appearance of singular points of curvature. Moreover, spacetime will inevitably lose its physical meaning at the singularity.
To solve this problem, scientists have proposed various schemes to avoid the occurrence of singularities. In 1933, Bardeen proposed a possible idea \cite {15}, that is, replacing the metric function $g^{rr}=1-\frac{2M}{r}$, which yielded a singularity with a function in the form of $g^{rr}=1-\frac{2Mr^2}{(r^2+Q^2)^{3/2}}$.  Therefore, the curvature tensor does not diverge when $r\rightarrow0$. In general, a black hole with this metric is referred to as a regular  black hole. Subsequently, it was found that the metric \cite{9} can be derived from a nonlinear electromagnetic field with magnetic charge; thus, the regular black hole can be regarded as a black hole with a nonlinear electromagnetic field magnetic charge. Consequently , various regular black holes \cite {10} have been proposed, and their thermodynamic properties \cite {11} and QNMs properties \cite {14} have been studied.


With the theory of gravity occupying an increasingly important position in astronomy and cosmology, particularly the direct detection of gravitational waves \cite {12} and the event \cite {13} wherein the humans first acquired photos of black holes \cite {12} in the past decade, general relativity has become the dominant basic theory in astronomy and cosmology. However,  certain astronomical phenomena cannot be perfectly explained via general relativity, such as dark energy and dark matter puzzles, and the physical phenomena in the period of quantum cosmology.
Thus, certain scientists assume that general relativity cannot perfectly express the gravity theory; therefore, various assumptions have been introduced to modify general relativity.
Various modified gravity theories have been formed, among which the high-dimensional and Gauss-Bonnet theories \cite{15} are two famous gravity theories, which hold  that the spacetime dimension of the real world is more than 3 + 1. Further, the spacetime dimension should contain additional dimensions that cannot be detected by existing means. This view is accepted by the quantum gravity theory represented by the string theory. The Gauss-Bonnet gravity introduces a Gauss-Bonnet term in the action to replace the Ricci scalar curvature of general relativity, which induces several interesting properties.A study \cite{16} derived and studied a static regular Bardeen black hole of a 5-dimensional Gauss-Bonnet. A black hole can successfully avoid the spacetime singularity of the black hole center and can smoothly return to the spacetime situation of the uncharged 5-dimensional Schwarzschild black hole; therefore, it has research value.


This study addressed the scalar field quasinormal ringing behavior of such black holes.
If an unstable black hole is provided a small perturbation, it generates an increasing number of vibrations under the perturbation and eventually collapses. Therefore, studying black hole perturbations can indicate the stability of the black hole; if a small perturbation is applied, the black hole will vibrate accordingly.
This vibration is divided into three stages. In the first stage (initial perturbation stage), the vibration is related to many factors, such as the nature of the perturbation itself and the nature of the black hole spacetime. However, in the subsequent two stages, the vibration of the stable black hole is only related to the nature of the black hole itself. The second stage is the quasinormal ringing stage, because the vibration waveform of this stage can be expressed using a complex function $\omega$. The real and imaginary parts of the function represents the vibration frequency and attenuation rate of the vibration amplitude, respectively .
The third stage can usually be described via a power-law trailing waveform. In the quasinormal ringing stage, the waveform is very obvious and the amplitude is relatively large; thus. it can well reflect the nature of the black hole itself. Therefore, the QNMs are considered to be the characteristic sound of the black hole itself. The study of QNMs is conducive to the identification and analysis of the properties of black holes in astronomy; thus, it has great physical significance.


The remainder of this paper is organized as follows. Section 2 describes the derivation process of the QNMs master equation of the 5-dimensional Gauss-Bonnet Bardeen regular black hole scalar field in spacetime. Section 3 employs the 6rd-order Padé  approximation method and the finite difference method to study the QNMs of scalar field perturbation in a black hole background. Finally, the discussions and conclusions are presented in Section 4.

\section{Scalar Perturbation of Bardeen Black Hole in 5-dimensional Gauss-Bonnet Gravity}
\renewcommand{\theequation}{2.\arabic{equation}} \setcounter{equation}{0}

A nonlinear electromagnetic field is introduced into 5-dimensional Gauss-Bonnet Gravity with action \cite{151}.

\bqn
\lb{action1}
S&=&\frac{1}{2}\int {\rm d}^5x\sqrt{-g}[R-2\Lambda+\alpha\mathcal{L}_{\mathcal{GB}}-\mathcal{L}(F)],\nb\\
\mathcal{L}_{\mathcal{GB}}&=&R_{\mu\nu\lambda\gamma}R^{\mu\nu\lambda\gamma}-4R_{\mu\nu}R^{\mu\nu}+R^2,
\eqn
where $R$, $\mathcal{L}_{\mathcal{GB}}$,  and $\alpha$ are the Ricci scalar curvature, Gaussian-Bonnet term, and GB coupling coefficient, respectively.

 The function $\mathcal{L}(F)$ represents the coupling term between gravity and nonlinear electrodynamics \cite{Cai_2021}.
 $\mathcal{L}(F)$ is an arbitrary function of the variable $F=\frac{1}{4}F_{\mu\nu}F^{\mu\nu}$, where $F_{\mu\nu}$ represents the Maxwell field-strength tensor.
\bqn
\lb{action2}
F_{\mu\nu}=\partial_{\mu}A_\nu-\partial_\nu A_\mu,
\eqn

 We investigate the Lagrangian term representing nonlinear electrodynamics in black hole systems using the Plebánski tensor $P_{\mu\nu}=F_{\mu\nu}\mathcal{L}_F$, where $\mathcal{L}_F=\frac{\partial \mathcal{L}}{\partial F}$. The nonlinear electrodynamics term of the considered system can be obtained through a Legendre transformation. Therefore, the Lagrangian density for nonlinear electrodynamics can be expressed as:
\bqn
\label{lagrangian}
\mathcal{L}=2P\mathcal{H}_P-\mathcal{H}
\eqn

Here, $P=\frac{1}{4}P_{\mu\nu}P^{\mu\nu}$ denotes the antisymmetric field, and $F_{\mu\nu}=\mathcal{H}_P P_{\mu\nu}$, where $\mathcal{H}_P=\frac{\partial \mathcal{H}}{\partial P}$. The structure function $\mathcal{H}(P)$ is provided by reference \cite{7}.
\bqn
\label{HP}
\mathcal{H}(P)=\frac{3}{2 s Q^4}\left(\frac{\sqrt{-2 Q^2 P}}{1+\sqrt{-2 Q^2 P}}\right)^{7 / 3},
\eqn



Where $s$ is a free parameter, and the relationship between $s$ and the magnetic monopole charge $Q$ is given by $s=\frac{|Q|}{2M}$, where $M$ represents the dimensionless mass of the celestial body.

The field equations can be obtained by extremizing the action \eqref{action1} with respect to $g_{\mu\nu}$ and $A_{\mu\nu}$ \cite{Cai_2021}.
\bqn
\label{ll1}
\begin{aligned}
   G_\mu^v&=2\left[\mathcal{H}_P P_{\mu \lambda} P^{v \lambda}-\delta_\mu^v\left(2 P \mathcal{H}_P-\mathcal{H}(P)\right)\right] \\
\end{aligned}
\eqn
\bqn
\label{fe1}
 \nabla_\mu P^{\mu\nu}=\frac{1}{\sqrt{-g}}\partial_\mu(\sqrt{-g}P^{\mu\nu})=0
\eqn

Next, considering the case of nonlinear electromagnetic fields, we constrain the electric field to be
\bqn
F_{\mu\nu}=E(r)(\delta_\mu^t\delta_\nu ^r-\delta_\nu^t \delta_\mu ^r)
\eqn

Where $F_{\mu\nu}$ represents the electromagnetic tensor as mentioned earlier. Assuming the black hole to be spherically symmetric, the determinant of the metric tensor is given by $\sqrt{-g}=r^2$.

Therefore, the field equation \eqref{fe1} can be equivalently expressed as $\frac{1}{r^2}\partial_{\mu}(r^2F^{\mu\nu}\mathcal{L}_F)=0$, implying $r^2P^{\mu\nu}=Constant$. Letting this constant be the charge $e$, we have

\bqn
\label{ll2}
P^{\mu\nu}=-\frac{Q}{r^2}(\delta^t_\mu\delta^r_\nu-\delta^t_\nu\delta^r_\mu)
\eqn

From \cite{Cai_2021}, setting $P=-\frac{Q^2}{2r^4}$ and substituting $P$ into \eqref{HP}, we can obtain the expression for $\mathcal{H}$:

\begin{equation}
    \mathcal{H}=\frac{3}{2se^4}\left(\frac{Q^2}{Q^2+r^2}\right)^\frac{7}{3}
\end{equation}

Then, based on \eqref{lagrangian}, utilizing Mathematica for computation yields the expression for $\mathcal{L}(r)$.
\begin{equation}
   \mathcal{L}(r)= \frac{r^4 \left(\frac{Q^2}{r^2}+4\right)-3 Q^4}{2 r^8 s \sqrt[3]{\frac{r^2}{Q^2}+1} \left(\frac{Q^2}{r^2}+1\right)^4}
\end{equation}

 In this theory, we consider the following 5-dimensional static spherically symmetric metric:
\begin{equation}
\begin{aligned}
ds^2=-f(r)dt^2+\frac{1}{f(r)}dr^2+r^2(d\theta^2+\\ \sin^2\theta 
d\phi^2+\sin^2\theta\sin^2\phi d\psi^2),
\end{aligned}
\end{equation}

According to the paper \cite{7}, the metric is given by $f(r)$:

 \begin{equation}
 \label{metrix1}
 f(r)=1+\frac{r^2}{4 \alpha}\left(1 \pm \sqrt{1+\frac{8 \alpha M}{\left(r^3+Q^3\right)^{4 / 3}}-\frac{8 \alpha}{l^2}}\right) .
 \end{equation}

Since we are discussing the case of the Bardeen black hole in asymptotically flat spacetime, we set the Planck length $l$ in $f(r)$ to approach infinity, implying that the cosmological constant tends to zero. Additionally, we adopt the "-" signature. Thus, we have the following metric:
\begin{equation}
\label{metrix}
f(r)=1+\frac{r^2}{4 \alpha}\left(1 - \sqrt{1+\frac{8 \alpha M}{\left(r^3+Q^3\right)^{4 / 3}}}\right) .
\end{equation}

Observing the principle of least action represented by \eqref{action1}, we understand that the Gauss-Bonnet (GB) term is a coupling term with $\alpha$ being the coupling coefficient. For the 5D Bardeen black hole with GB term and charge term, when $\alpha\rightarrow0$ and $Q\rightarrow0$, the black hole solution degenerates to the 5D Schwarzschild black hole case:
\begin{equation}
    f(r)\rightarrow 1-\frac{M}{r^2}
\end{equation}

\begin{figure*}
        \center
        \scriptsize
        \begin{tabular}{cc}
                \includegraphics[scale=0.55]{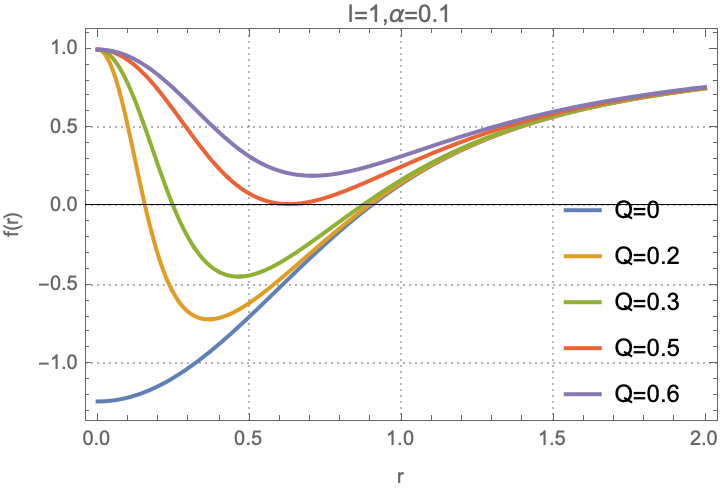} &    \includegraphics[scale=0.55]{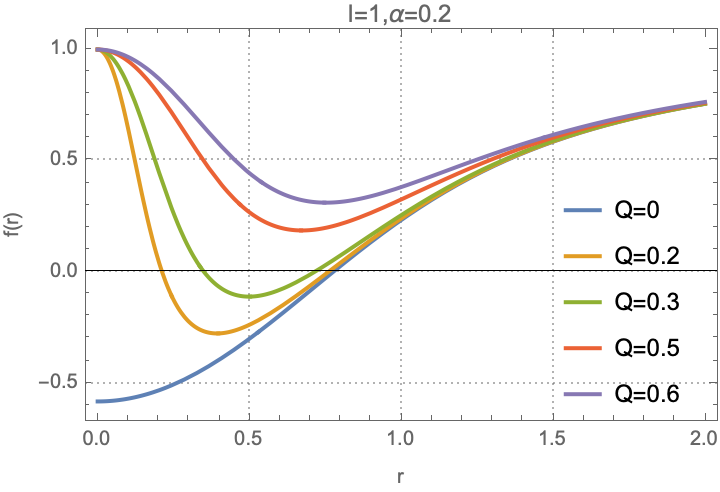}   \\
                (a) & (b) \\
        \end{tabular}
        \caption{$f(r)$ vs $r$ corresponding to GB parameter  $l=1,\alpha=0.1$ (a) and $ l=1,\alpha=0.2$ (b) for different values of charge $Q$}
        \label{f(r)}
        \vspace{-0.5em}
\end{figure*}

Here, $M$ represents the parameter mass of the black hole, as depicted in Figure \eqref{f(r)}. Clearly, when the magnetic charge $Q\neq0$, the Bardeen black hole does not possess singularities.

\begin{figure*}
        \center
        \scriptsize
        \begin{tabular}{cc}
                \includegraphics[scale=0.55]{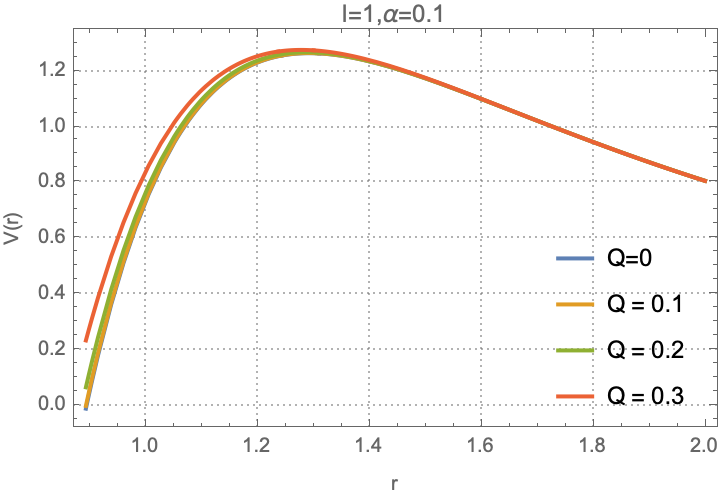} &    \includegraphics[scale=0.55]{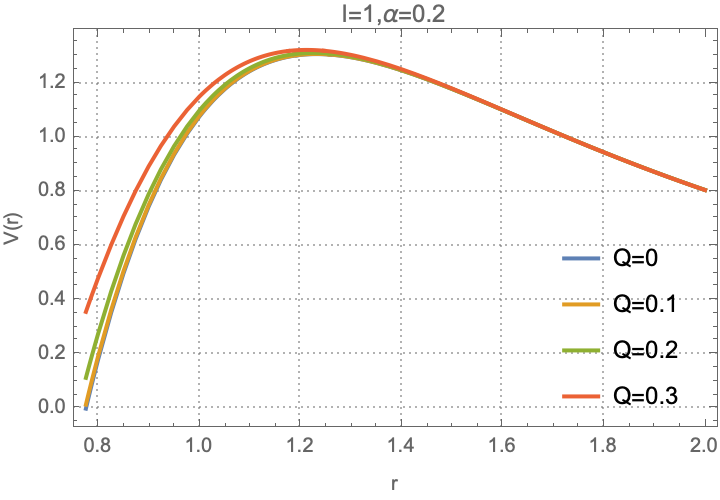}        \\
                (a) & (b) \\
        \end{tabular}
            \caption{  $V(r)$ vs $r$ corresponding to   $l=1,\alpha=0.1$ (a) and $ l=1,\alpha=0.2$ (b) for different values of charge $Q$}
        \label{V}
        \vspace{-0.5em}
\end{figure*}

In a curved spacetime, massless scalar field perturbations can be described by the Klein-Gordon equation:
\begin{equation}
    \begin{aligned}
       \nabla^\nu\nabla_\nu\Phi=0
    \end{aligned}
\end{equation}
or
\begin{equation}
    \frac{1}{\sqrt{-g}}\frac{\partial}{\partial x^\mu}\left(\sqrt{-g}g^{\mu\nu}\frac{\partial}{\partial x^\nu}\Phi\right)=0.
\end{equation}
In the 5-dimensional Bardeen black hole spacetime in Gauss-Bonnet gravity, the above field equation becomes
\begin{equation}
\begin{aligned}
    &-\frac{1}{f(r)}\frac{\partial^2\Phi}{\partial t^2}+\frac{1}{r^3}\left[\left(3r^2f(r)+r^3\frac{{\rm d} f(r)}{{\rm d} r}\right)\frac{\partial \Phi}{\partial r}\right.\\
    &\left.+r^3 f(r)\frac{\partial^2\Phi}{\partial t^2}\right]+\frac{1}{r^2}\left(2\cot\theta\frac{\Phi}{\partial\theta}+\frac{\partial^2\Phi}{\partial\theta^2}\right)+\frac{1}{r^2\sin^2\theta}\\
    &\times\left(\cot\phi\frac{\partial\Phi}{\partial\phi}+\frac{\partial^2\Phi}{\partial\phi^2}\right)+\frac{1}{r^2\sin^2\theta\sin^2\phi}\frac{\partial^2\Phi}{\partial\psi^2}=0    
\end{aligned}
\end{equation}
Because  the Bardeen black hole is spherically symmetric, we can solve the above equation by separating the variables. Therefore, function can be decomposed into 
\begin{equation}
    \begin{aligned}
        \Phi(t,r,\theta,\phi,\psi)=\sum_{l,m,k} r^{-\frac{3}{2}} \Psi(t,r)Y_{lmk}(\theta,\phi,\psi)
    \end{aligned}
\end{equation}
where $Y_{lmk}(\theta,\phi,\psi)$ is the angle term, and is the extra dimension term. As study \cite{new2} showed that the radial part of a scalar field can be written as
\bqn
\lb{fun1}
\left(\frac{\partial^2}{\partial t^2}-\frac{\partial^2}{\partial r_*^2}+V(r)  \right) \Psi(t,r_*)=0,
\eqn
$r_*=\int\frac{1}{f(r)}dr$ is the tortoise coordinate and $V(r)$ is the potential function\cite{Moura_2021,Daghigh_2007}.
\bqn
\lb{potential}
V(r)&=&f(r)\left[\frac{l(l+2)}{r^2}+\frac{3f'(r)}{2r}+\frac{3f(r)}{4r^2}\right]\nb\\
&=&f(r)\cdot\left[\ \frac{3+4l(2+l)}{4r^2}+\frac{6r^3(Q^3+r^3)^{-\frac{7}{3}}}{\sqrt{1+\frac{8\alpha}{(Q^3+r^3)^\frac{4}{3}}}}\right. \nb\\
&&\left.+\frac{15}{16\alpha}\left(1-\sqrt{1+\frac{8\alpha}{(Q^3+r^3)^\frac{4}{3}}}\right)\right],
\eqn
where $l=0,1,2,\cdots$ is the angular quantum number. The potential function for this perturbation is shown in Figure\eqref{V}.

Let $\Psi(t,r)=Q^{-i\omega t}R(r)$; then, the above equation can be reduced to the standard  master equation of QNMs in static black hole spacetime \cite{2}:
\bqn
\lb{MasterEqu1}
\frac{{\rm d}^2}{{\rm d}r_*^2}R(r)+\left[\omega^2-V(r)\right]R(r)=0
\eqn
As the potential function vanishes $V\rightarrow0$ at the event horizon $r=r_p$ and at infinity $r\rightarrow\infty$, the above equation can be reduced to a standard wave function form.
\bqn
\lb{MasterEqu2}
\frac{{\rm d}^2}{{\rm d}r_*^2}R(r)+\omega^2R(r)=0
\eqn

Therefore, the analytical solution $R=\exp\left(\pm i\omega r_*\right)$ can be obtained at the event horizon and infinity, where $+$ and $-$ represent the waves propagating along the $r_*$ direction and against the $r_*$ direction, respectively. Further, the boundary conditions of  QNMs require only incident waves propagating along the direction opposite to that of $r_*$ at the horizon. This is because the classical black hole is a star that only allows particles to enter and not come out. At infinity, only the outgoing wave propagating along the direction of $r_*$ is assumed. Using the master equation and boundary conditions, we can study the QNMs under the scalar field perturbation of the black hole.

\section{Numerical Results of QNMs}
\renewcommand{\theequation}{3.\arabic{equation}} \setcounter{equation}{0}

Although the black hole QNMs equation is not that complicated, directly obtaining an analytical solution is challenging. Therefore, various numerical methods have been proposed for studying the QNMs of black holes, such as the WKB approximation \cite{30}, continuous fraction method \cite{leaver}, finite-difference method \cite{time}, asymptotic iteration method \cite{33}, and matrix method \cite{34}. In this Section, the  QNMs of a black hole are examined using the 6rd-order WKB and Padé  approximation and the finite difference method.

\subsection{Using  6th order Padé  approximation  to calculate Quasinorml Modes frequency.}

The WKB method is a semi-classical approximation method for solving the Schrodinger equation proposed by Wenzel, Kramers, and Brillouin \cite{wkb1,wkb2,wkb3}. In quantum mechanics, the WKB approximation is often used to address the problem of a particle penetration barrier. In 1985, Schutz and Will first used the WKB method to calculate the QNM frequency \cite{4} of black holes. To  improve its accuracy, physicists have developed it to the 3rd-, 6th-, and 13th-orders \cite{5}.





The  6th order WKB formula is as follows\cite{Ay_n_Beato_1998},
\bqn
\lb{WKB1}
\omega^2&=&[V_0+(-2V_2)^{\frac{1}{2}}\Lambda]\nb\\
&&-i\left(N+\frac{1}{2}\right)(-2V_2)^{\frac{1}{2}}(1+\Omega)-\Gamma
\eqn
where,
\bqn
\lb{WKB2}
\alpha&=&N+\frac{1}{2},\quad N=0,1,2\cdots \nb\\
V_n&=&f(r)\cdot\left.\frac{{\rm d}V_{(n-1)} }{{\rm d} r}\right|_{r=r(r_0)}\nb \\
\Gamma&=&i \sqrt{-2 V_2} \left(\Lambda_4+\Lambda_5+\Lambda_6\right);
\eqn
where  $\Lambda_4,\Lambda_5,\Lambda_6$ can be found in \cite{Ay_n_Beato_1998}, $V_0$ is the maximum value of the effective potential function. $r_0$ is determined from $V'(r)=0$. $n_{max}=12$. In Eq.\eqref{WKB1}, because $\omega\in \mathbb{Z}$, $\omega$ contains both the real and imaginary parts, $\Lambda$ and $\Omega$ can be expressed as follows,

\bqn
\lb{WKB3}
\Lambda&=&\frac{1}{  \left(-2V_2\right)^\frac{1}{2} }\Bigg\{\frac{1}{8}\left(\frac{V_4   }{V_2   }\right)\left(\frac{1}{4}+\alpha^2\right)
\\&&-\frac{1}{288}\left(\frac{V_3 }{V_2 }\right)^2\left(7+60\alpha^2\right)\Bigg\} ,\nb\\
\Omega&=&-\frac{1}{2V_2 }\Bigg\{\frac{5}{6912}\left(\frac{V_3}{V_2}\right)^4(77+188\alpha^2)-\frac{1}{384}\cdot\nb\\&&\left(\frac{(V_3)^2V_4}{(V_2)^3 }\right)\left(51+100\alpha^2\right)
+ \frac{1}{2304}\left(\frac{V_4}{V_2}\right)^2\cdot\nb\\&&\left(67+68\alpha^2\right)+\frac{1}{288}\left(\frac{V_3V_5}{(V_2)^2}\right)\left(19+28\alpha^2\right)\nb\\&&-\frac{1}{288}\frac{V_6}{V_2}\left(5+4\alpha^2\right)   \Bigg\}\nb\\
\eqn


As indicated in reference \cite{Konoplya_2003}, when the angular quantum number $l$ and the energy level $n$ are close, the third-order WKB method exhibits significant errors. For instance, when $l=N=0$, the error of the third-order WKB method reaches $10\%$. Moreover, with increasing dimensions, such errors continue to amplify. To enhance the accuracy of the WKB formula further, this paper will employ the 6th order Padé approximation method  for computation.

Next, we introduce the construction of the  6th order Padé approximation polynomial. Expressing \eqref{WKB1} in polynomial form yields \cite{Konoplya_2019},
\begin{equation}
    \label{dxs}
    \begin{aligned}
\omega^2&=V_0+A_2(\alpha^2)+A_4(\alpha^2)+A_6(\alpha^2)+\cdots\\ 
        &-i \alpha\sqrt{-2 V_2}(1+A_3(\alpha^2)+A_5(\alpha^2)+\cdots ) 
    \end{aligned}
\end{equation}
where $A_k(\alpha^2)$ is a polynomial dependent on $V_2,V_3,\cdots,V_{2k}$ and $\alpha^2$. $k$ represents the order of the Padé approximation. This order corresponds to the order in the WKB method.

We define new polynomials $P_k(\epsilon)$ by introducing multiple powers of a new parameter $\epsilon$ into the right-hand side of \eqref{dxs} \cite{Konoplya_2019}:
\begin{equation}
\begin{aligned}
        P_{k}(\epsilon)&=V_0+A_2(\alpha^2)\epsilon^2+A_4(\alpha^2)\epsilon^4+\cdots\\ 
        &-i \alpha\sqrt{-2 V_2}(\epsilon+A_3(\alpha^2)\epsilon^3+A_5(\alpha^2)\epsilon^5+\cdots ) 
\end{aligned}
\end{equation}

Subsequently, we construct a family of rational functions,
\begin{equation}
\label{pade1}
    P_{\hat{n}\backslash \hat{m}}(\epsilon)=\frac{Q_0+Q_1\epsilon+\cdots+Q_{\hat{n}} \epsilon^{\hat{n}}}{R_0+R_1\epsilon+\cdots+R_{\hat{m}} \epsilon^{\hat{m}}}
\end{equation}
In this case, \eqref{pade1} has the following equivalent relationship,
\begin{equation}
    \label{djgx}
 P_{\hat{n}\backslash \hat{m}}(\epsilon)-P_k(\epsilon)=\mathcal{O}(e^{k+1})
\end{equation}

The calculation method for $Q_i\cdots Q_n$ and $R_0\cdots R_n$ is similar to that of $A_n$. Similarly, when $\epsilon=0$, we have,
\begin{equation}
    \omega^2= P_{\hat{n}\backslash \hat{m}}(1)
\end{equation}
The k-th order WKB formula can be transformed into an alternative form corresponding to $\hat{m},\hat{n}=1,2,3\cdots$.



\begin{figure*}
        \center
        \scriptsize
        \begin{tabular}{cc}
                \includegraphics[scale=0.5]{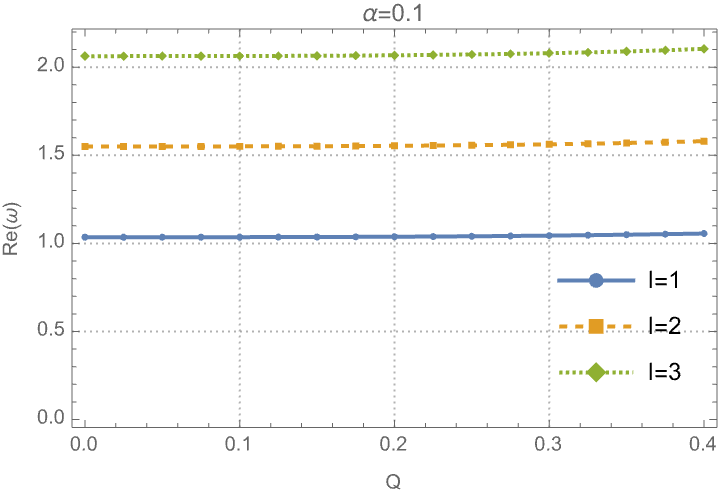} & 
                
                \includegraphics[scale=0.5]{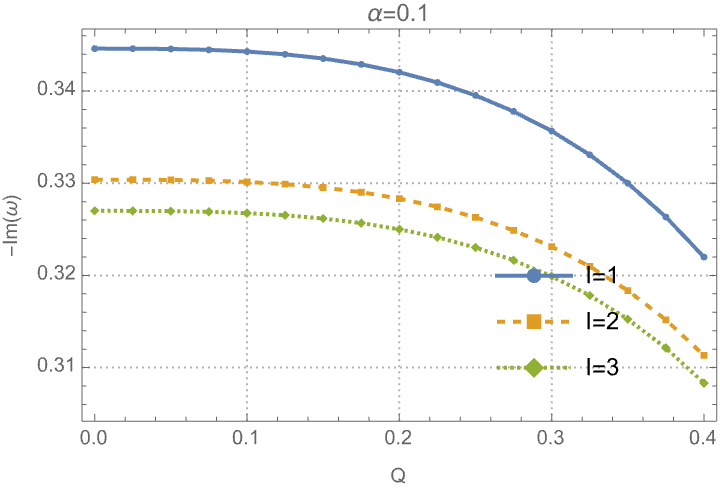}        \\
                (a) & (b) \\
                \includegraphics[scale=0.5]{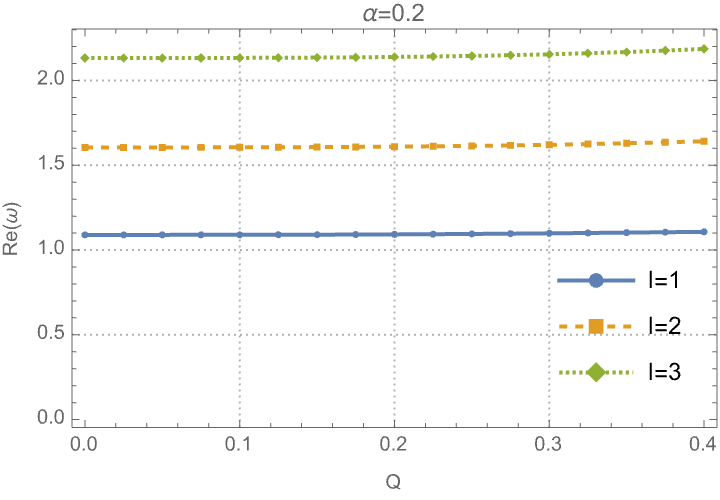} &    \includegraphics[scale=0.5]{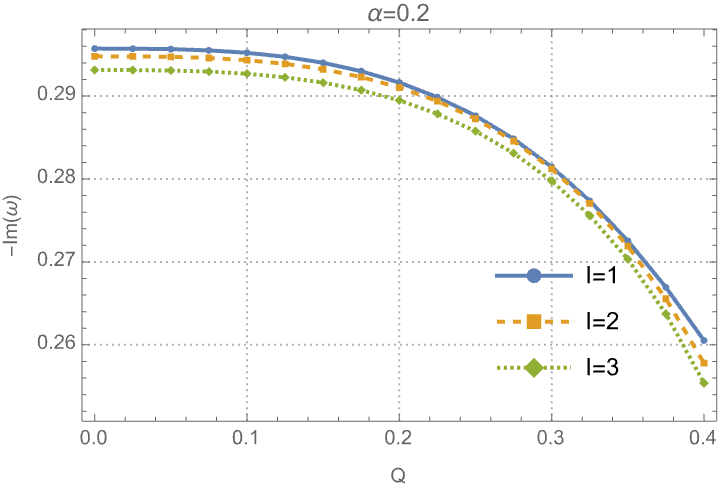}          \\
                (c) & (d) \\
        \end{tabular}
         \caption{Bardeen BH QNMs frequency vs $Q$ under scalar field perturbation in 5-dimensions for $M = 1$ with 6-th order WKB, (a)(b) for $\alpha = 0.1$, (c)(d) for $\alpha = 0.2$. (a)(c) is the real part of QNMs. (b)(c) is the imaginary part of QNMs.}
        \label{omega}
        \vspace{-0.5em}
\end{figure*}

Using the Padé approximation method, we first studied the variation trend of the  QNMs of black holes under the condition that the Gauss-Bonnet parameter $\alpha$ is fixed(Many thanks to Konoplya for the mathematica code here\cite{code}). Figure\eqref{omega} knows that it is not difficult to find that fixed $\alpha$. In the case of scalar field perturbation, as Q increases, the real part of the QNM frequency will gradually increase, and the absolute value of its imaginary part of the QNM frequency will gradually decrease. Further, vertical comparison shows that $\alpha$ has minimal effect on the real part of the QNMs, and that it significantly affects the imaginary part of the QNMs.

\begin{figure*}
        \center
        \scriptsize
        \begin{tabular}{cc}
                \includegraphics[scale=0.5]{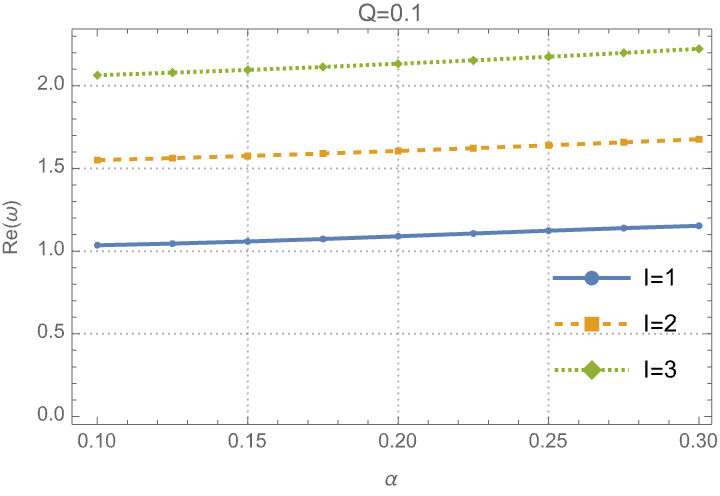} &    \includegraphics[scale=0.5]{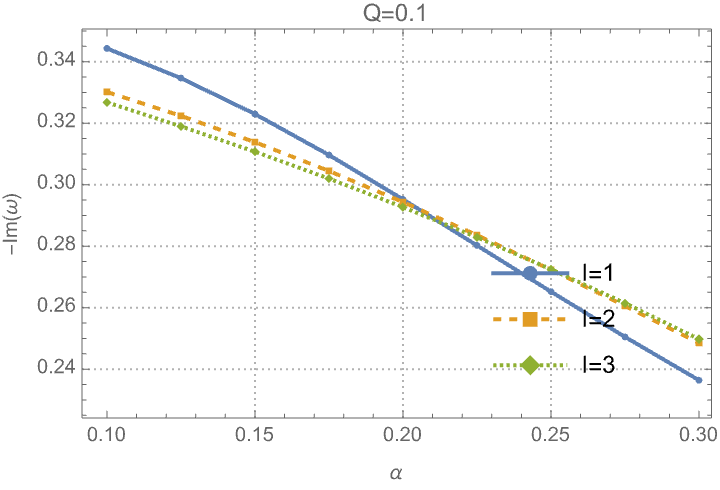}        \\
                (a) & (b) \\
                \includegraphics[scale=0.5]{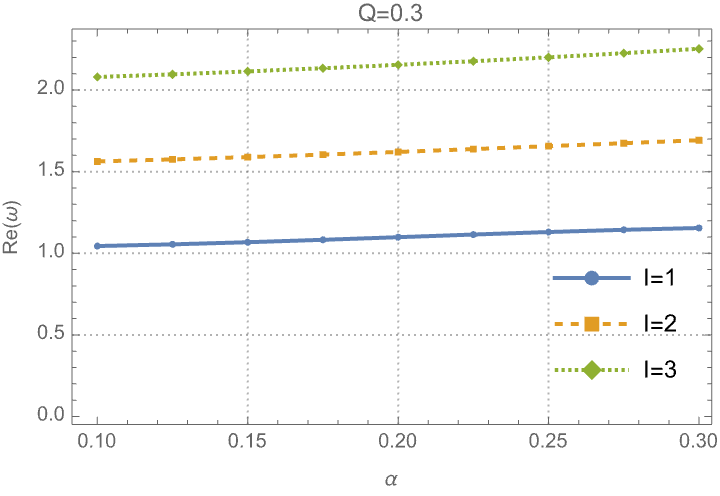} &    \includegraphics[scale=0.5]{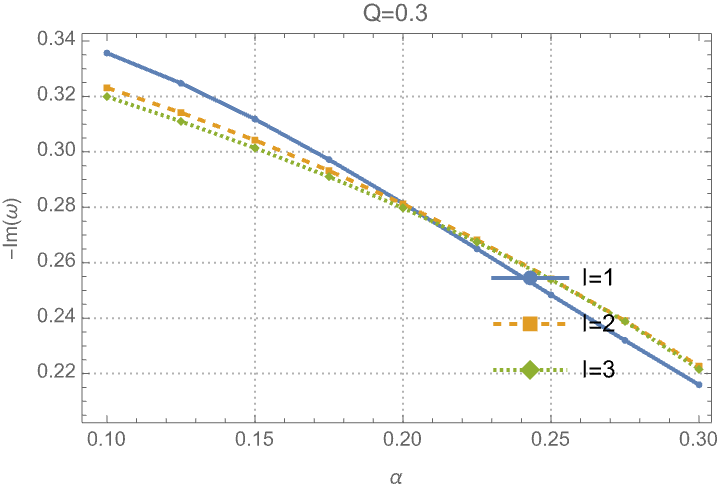}        \\
                (c) & (d) \\
        \end{tabular}

         \caption{Bardeen BH QNMs frequency vs $\alpha$ under scalar field perturbation in 5-dimensions for $M = 1$, (a)(b) for $Q = 0.1$, (c)(d) for $Q = 0.3$. (a)(c) is the real part of QNMs. (b)(c) is the imaginary part of QNMs.}
        \label{omega52}
        \vspace{-0.5em}
\end{figure*}

We also studied the change in QNMs of the black hole owing to the Gauss-Bonnet parameter $\alpha$ when the parameter $Q$ is fixed. Figure \eqref{omega52} shows that to determine that for fixed Q, in the case of scalar field perturbation, with increase in $\alpha$, the real part of the QNMs frequency gradually increases, and the absolute value of the imaginary part of the QNMs will gradually decrease.

Observing the plot, we notice an intriguing phenomenon. With $Q$ fixed, when $\alpha\approx 0.21$, the quasi-normal modes for $l=1$ abruptly decrease, causing them to fall below the cases for $l=2$ and $l=3$. This suggests that, under fixed $Q$, there isn't a strictly linear relationship between the quasi-normal modes of black holes and the coupling coefficient $\alpha$ of the Gauss-Bonnet term. There exists some degree of nonlinearity between the two.

A comparison of Figures \eqref{omega} and \eqref{omega52} revealed that the effect of $\alpha$ on the QNMs of black holes is greater than that of $Q$ on black holes.

\begin{figure*}
        \center
        \scriptsize
        \begin{tabular}{cc}
                \includegraphics[scale=0.6]{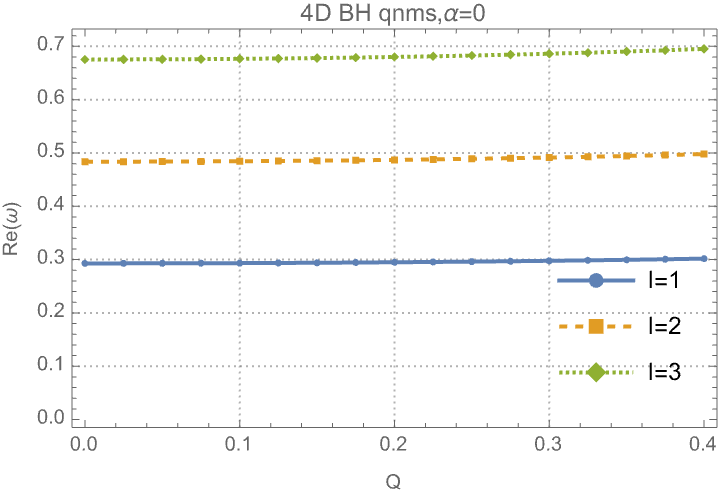} &    \includegraphics[scale=0.6]{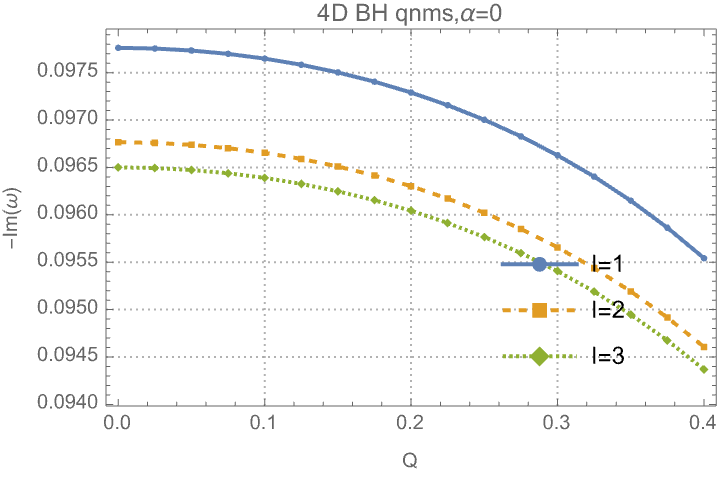}        \\
                (a) & (b) \\
                \includegraphics[scale=0.6]{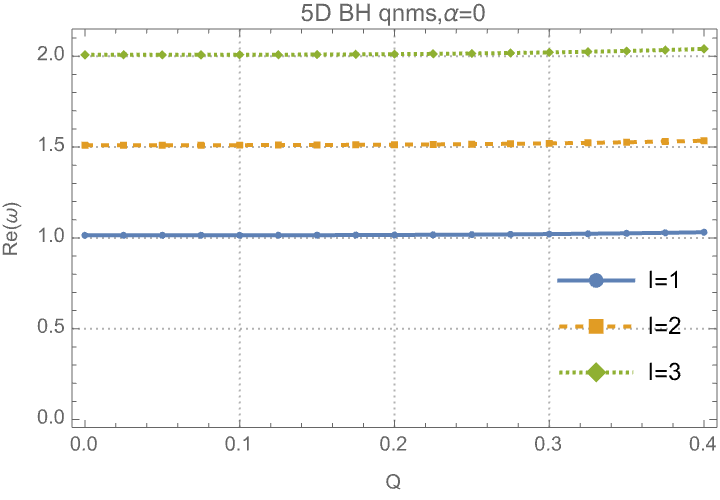} &    \includegraphics[scale=0.6]{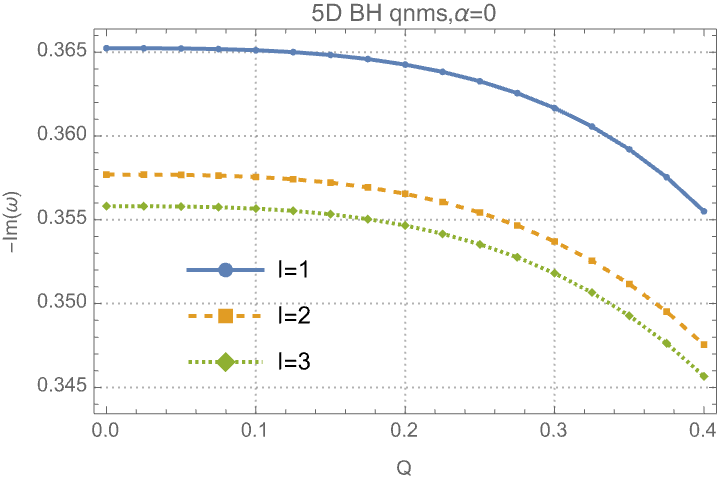}        \\
                (c) & (d) \\
        \end{tabular}
        \caption{The relationship between the quasi-normal mode frequencies and $Q$ for a Bardeen black hole with $M = 1$ and $\alpha=0$, under scalar field perturbations in four and five dimensions, is depicted. Panels (a) and (c) show the real parts of the quasi-normal mode frequencies, while panels (b) and (d) illustrate the imaginary parts.}
        \label{omega4}
        \vspace{-0.5em}
\end{figure*}



Since the Einstein-Gauss-Bonnet (EGB) gravity is introduced in a five-dimensional space, it decouples when $D=4$. According to \cite{4dbardeen}, the metric of a Bardeen black hole in four-dimensional AdS spacetime is given by:
\begin{equation}
\label{bardeen4}
f(r)=1-\frac{2 M r^2}{\left(r^2+Q^2\right)^{\frac{3}{2}}}-\frac{\Lambda r^2}{3}
\end{equation}
Similarly, since we are studying the case of Bardeen black holes in asymptotically flat spacetime, we set $\Lambda=0$. In this case, the Bardeen black hole is free of singularities. Its metric takes the following form:
\begin{equation}
\label{bardeen5}
f(r)=1-\frac{2 M r^2}{\left(r^2+Q^2\right)^{\frac{3}{2}}}
\end{equation}

Similar to the 5D case, when $Q\rightarrow 0$, the metric of the Bardeen black hole degenerates to that of the Schwarzschild black hole:
\begin{equation}
    f(r)\rightarrow1-\frac{2M}{r}
\end{equation}

The quasi-normal modes of Bardeen black holes in 4D asymptotically flat spacetime are depicted in \eqref{omega4}. Comparing with Bardeen black holes coupled with non-Gauss-Bonnet terms in five dimensions, it is evident that the real and imaginary parts of the quasi-normal modes vary significantly in higher dimensions. This suggests that in higher dimensions, the frequency of gravitational wave emission from black holes will be higher, with a greater damping rate.


 The real and the absolute values of  imaginary parts of the QNMs frequency of the Bardeen black hole are larger than those of the Schwarzschild black hole in the same dimension. Moreover, the QNMs frequency and decay rate of Bardeen black holes are higher, and the comparison of Schwarzschild black holes and Bardeen black holes in 5-dimensional asymptotically flat spacetime is shown in Table \eqref{vs}.

\begin{table}[H]
    \centering
    \caption{The quasi-normal mode frequencies of 5D Schwarzschild black holes and Bardeen black holes, calculated using the 6th order Padé approximation with $M=1$, are obtained.}
    \begin{tabular}{|l|l|l|}
    \hline
        ~ & \makecell{$Q=0,\alpha=0$\\(Schwarzschild BH) }&\makecell{ $Q=0,\alpha=0.1$\\(Gauss-Bonnet BH)}\\ \hline
        $l=0$ & 0.491229- 0.411 i & 0.489581 - 0.353576 i \\ \hline
        $l=1$ & 1.00348 - 0.363473 i & 1.02654 - 0.326329 i \\ \hline
        $l=2$ & 1.50707 - 0.357872 i & 1.54643 - 0.325853 i \\ \hline
        $l=3$ & 2.0065 - 0.355947 i & 2.06074 - 0.325686 i \\ \hline
        ~ & \makecell{$Q=0.3,\alpha=0.1$\\(Bardeen BH)} & \makecell{$Q=\alpha=0.3$\\(Bardeen BH)} \\ \hline
        $l=0$ & 0.483916 - 0.342984 i & 0.496418 - 0.285891 i \\ \hline
        $l=1$ & 1.03311 - 0.318495 i & 1.10931 - 0.239491 i \\ \hline
        $l=2$ & 1.55822 - 0.318541 i & 1.68429 - 0.228853 i \\ \hline
        $l=3$ & 2.07713 - 0.318538 i & 2.25117 - 0.223586 i\\ \hline
    \end{tabular}
    \label{vs}
\end{table}


Both Bardeen black holes and Schwarzschild black holes exhibit a noticeable increase in the real part of the QNMs in 5-dimensional spacetimes($l=1,l=2,l=3$), while the imaginary part of the QNMs remains nearly constant. This constitutes an intriguing and thought-provoking phenomenon.


\subsection{Using the finite difference method to solve the Quasinormal ringing}

The finite-difference method can also be used to study Eq. (\eqref{fun1}) directly. To achieve this, we introduce the following coordinate transformations:
\begin{equation}
    \begin{aligned}
        u=t-r_*,v=t+r_*.
    \end{aligned}
    \label{daihuan}
\end{equation}
Consequently, Eq. (\eqref{fun1}) is reduced to
\begin{equation}
    \begin{aligned}
        \frac{\partial^2\Psi(u,v)}{\partial u\partial v}+\frac{1}{4}V(r(u,v))\Psi(u,v)=0
    \end{aligned}
\end{equation}
The above equation is then discretized in a finite-difference format \cite{6,8}, such that
\bqn
\label{chafen}
\Psi_N&=&\Psi_E+\Psi_W-\Psi_S\nb\\
&&-\Delta u\Delta v\cdot V(r)\frac{\Psi_W+\Psi_E}{8}
+{\cal O}(\Delta^4),
\eqn
where $S=(u,v),W=(u+\Delta u,v),E=(u,v+\Delta v),N=(u+\Delta u,v+\Delta v)$ Gaussian pulses were used as the initial conditions.
\begin{equation}
    \begin{aligned}
        \Psi(u=u_0,v)&=\exp\left[-\frac{(v-v_c)^2}{2\sigma^2}\right],\\
        \Psi(u,v=v_0)&=0.
    \end{aligned}
\end{equation}

\begin{figure*}[htbp!]
        \center
        \scriptsize
        \begin{tabular}{cc}
                \includegraphics[scale=0.36]{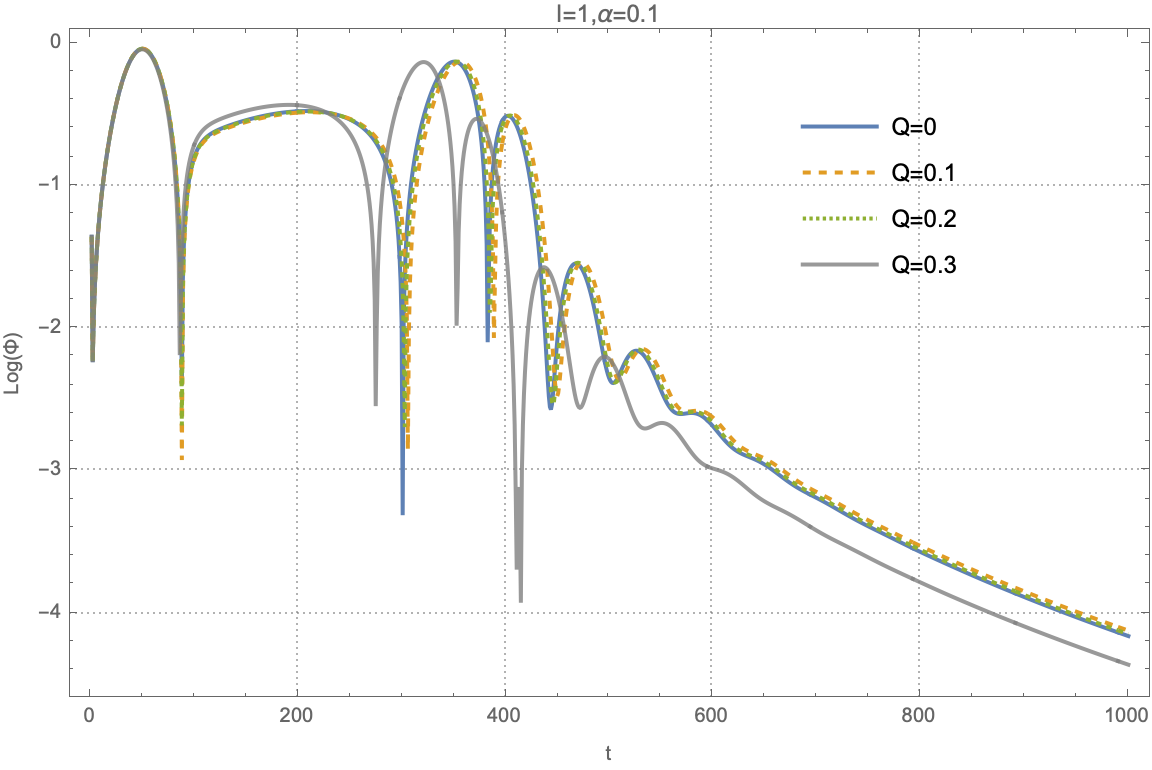} &    \includegraphics[scale=0.36]{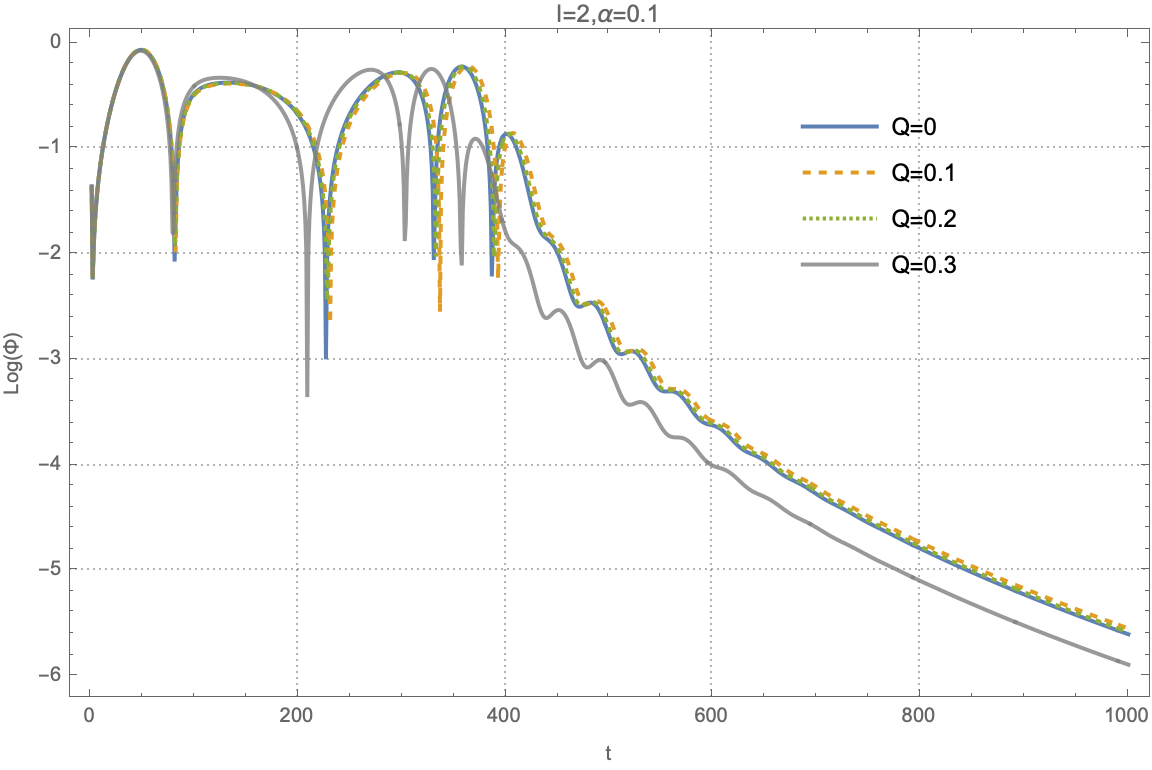}        \\
                (a) & (b) \\
                \includegraphics[scale=0.36]{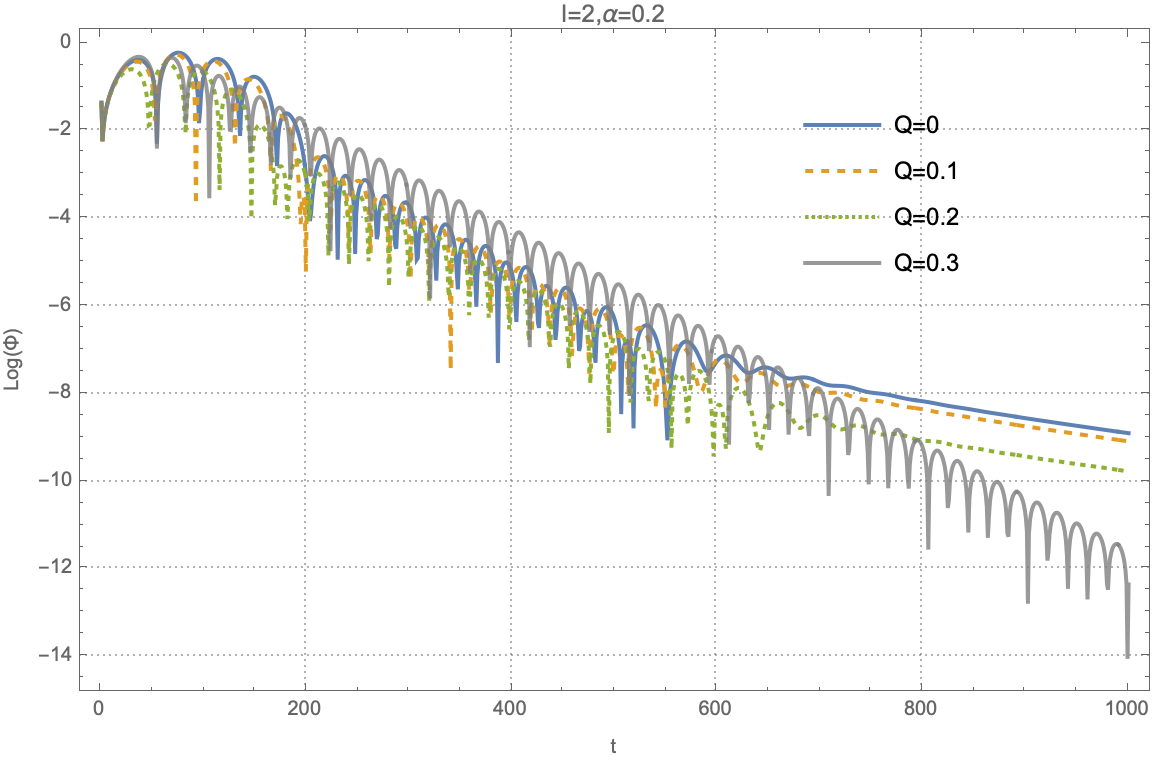} &    \includegraphics[scale=0.21]{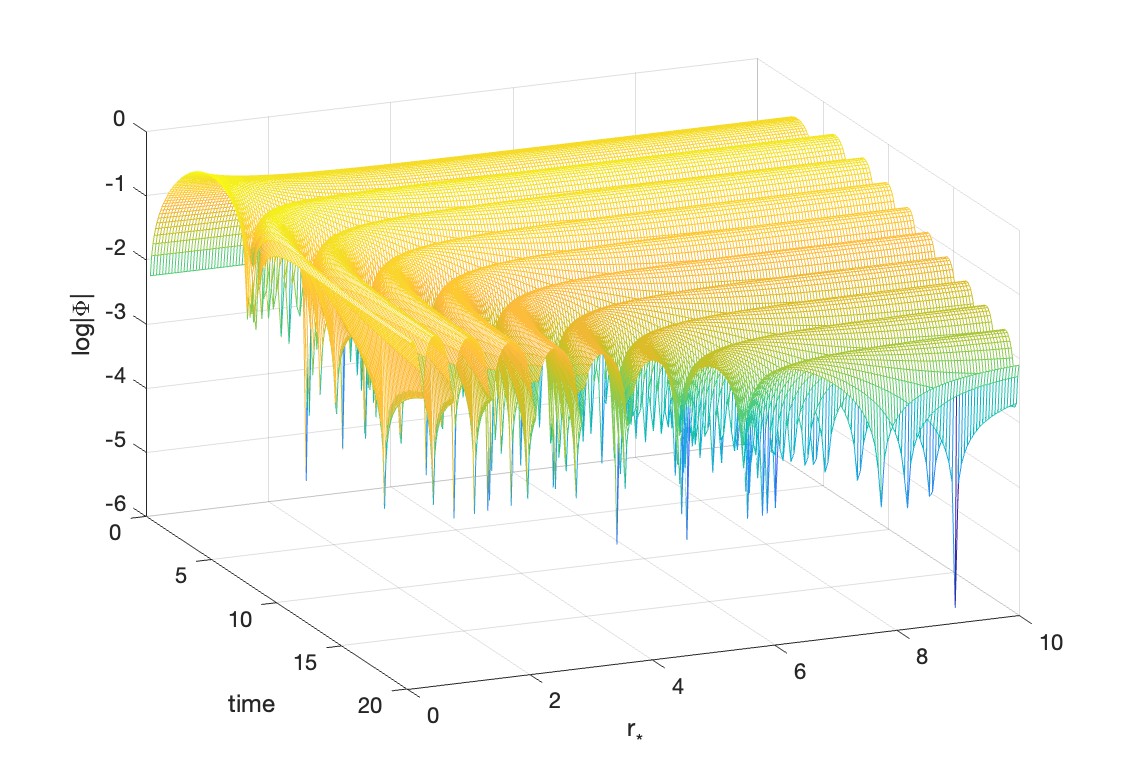}        \\
                (c) & (d) \\
        \end{tabular}
             \caption{Quasinormal ringing of the Bardeen black hole $l=1,\alpha=0.2$(a),$l=2,\alpha=0.2$(b),$l=2,\alpha=0.2$(c)(d) in 5-dimensional asymptotically flat spacetime}
        \label{ring}
        \vspace{-0.5em}
\end{figure*}
where $\sigma=5,v_c=2$. The variable of the potential function $V(r)$ in difference equation \eqref{chafen} is $r$ instead of $r_*$. We must compute $r_*$ from $u,v$ and then find $r$ by determining the inverse function of $r_*=\int \frac{1}{f(r)}{\rm d}r$.  Because of the complex nature of the metric \eqref{metrix} discussed, it is difficult to obtain an analytical solution for $r_*=r_*(r)$. Herein, we used the DNSolve command in Mathematica to solve the differential equation.

\begin{equation}
    \begin{aligned}
        \frac{{\rm d}r}{{\rm d}r_*}=\frac{1}{f(r)}
    \end{aligned}
\end{equation}

We assumed that the initial value of the equation is $$r(-50)=r_0+0.000000000000001,$$ $r_0$ is the event horizon radius of the Bardeen black hole. The initial value was set such that the value range of $r_*$ after the turtle coordinate transformation was $ (-\infty, \infty)$.We selected $r_*=-50$ as the starting point of $r_*$ and $r(r_*)=r_0+0.000000000000001$ as the starting point of r. As the final QNMs image only reflects the release of gravitational waves over a period of time, different choices of the initial value only shifts the ringing image left and right. Moreover, the actual frequency remains unaffected. The result of the finite difference method is calculated as shown in Figure\eqref{ring}.


Figure \eqref{ring} illustrates the time-domain variation of the scalar field perturbations for different $l,Q,\alpha$. where (a) $\thicksim$(c) shows only the dynamical behavior at $r_*=0$, and (d) shows the ringing in the full time domain. A comparison of (a) and (b) revealed that the QNMs frequency increased and the QNMs decay rate decreased with increase in $l$. This is consistent with the fact that as $l$ increases, $\Re(\omega)$ increases and $|\Im(\omega)|$ decreases, as illustrated in Figure \eqref{omega}(a)(b). A comparison of (b) and (c) revealed that the QNM frequency increased and decay rate decreased when the GB coupling coefficient $\alpha$ increased, which is consistent with the increase in $\alpha$, increase in $\Re(\omega)$, and decrease in $|\Im(\omega)|$, as shown in Figures \eqref{omega}(c) and (d). Simultaneously, observing the change in $Q$ in Figure\eqref{ring} revealed that an increase in the magnetic charge increased the QNM frequency of the black hole and slows down the decay rate.

\section{Conclusion}
\renewcommand{\theequation}{4.\arabic{equation}} \setcounter{equation}{0}

This study investigated the nature of the quasinormal ringing of a Bardeen black hole within a 5-dimensional asymptotically flat spacetime under a scalar field perturbation. We obtained the effective potential of the radial equation by solving the massless Klein-Gordon equation. We found that the increase in the magnetic charge $Q$ can increase its peak value, thus affecting the QNMs frequency of the black hole. An increase in the GB coupling coefficient $\alpha$ also results in a higher peak effective potential. We used Padé approximation to compute the seemingly regular frequency of this Bardeen black hole spacetime under scalar field perturbations.
The real part of the frequency increases with $l,\alpha,Q$ and the absolute value of the imaginary part of the frequency decreased with $l,\alpha,Q$. We addressed the problem using the finite difference method and plotted the dynamics of the scalar field of $\Phi$ as a function of time and radius. The quasinormal ringing results illustrated that the magnetic charge $Q$ accelerated the gravitational wave decay of the Bardeen black hole.
We also compared the Bardeen black holes in different dimensional spacetimes and found that the real and imaginary parts of the QNMs frequency increased with increase in the dimension.

\newpage
\bibliography{ref}

\end{document}